\documentclass{article}

\usepackage{arxiv}

\usepackage[utf8]{inputenc} % allow utf-8 input
\usepackage[T1]{fontenc}    % use 8-bit T1 fonts
\usepackage{hyperref}       % hyperlinks
\usepackage{url}            % simple URL typesetting
\usepackage{booktabs}       % professional-quality tables
\usepackage{amsfonts}       % blackboard math symbols
\usepackage{nicefrac}       % compact symbols for 1/2, etc.
\usepackage{microtype}      % microtypography
\usepackage{lipsum}		% Can be removed after putting your text content
\usepackage{graphicx}
\usepackage{amsmath}
\usepackage{amssymb}
\usepackage{color}

\title{The gravitational field of the SdS space-time}

\author{ H. M. Manjunatha$^{*}$ and S. K. Narasimhamurthy$^{**}$ \\
	Department of PG Studies and Research in Mathematics\\
	Kuvempu University, Jnana Sahyadri\\
	Shankaraghatta - 577 451 \\
    Shivamogga, Karnataka, India\\
	$^{*}$\texttt{manjunathahmnmt@gmail.com} \\
    $^{**}$\texttt{nmurthysk@gmail.com} \\
	\And
	Z. Nekouee \\
	Department of Mathematics \\
    Faculty of Mathematical Sciences \\
    University of Mazandaran \\
    P. O. Box 47416-95447, Babolsar, Iran \\
	\texttt{z.nekouee@stu.umz.ac.ir} \\
}

\date{}

\renewcommand{\headeright}{}
\renewcommand{\undertitle}{Preprint of an article published in [International Journal of Geometric Methods in Modern Physics, 17, 05, 2020, 2050069] [doi: 10.1142/S0219887820500693] \copyright [copyright World Scientific Publishing Company] [https://www.worldscientific.com/worldscinet/ijgmmp]}

\hypersetup{
pdftitle={The gravitational field of the SdS space-time},
pdfsubject={math.DG, gr-qc},
pdfauthor={H. M. Manjunatha, S. K. Narasimhamurthy, Z. Nekouee},
pdfkeywords={Gravitational field, SdS space-time, Einstein space, Gaussian curvature},
}

\begin{document}
\maketitle

\begin{abstract}
The goal of this paper is to study SdS space-time and its gravitational field with consideration of the canonical form and invariants of the curvature tensor. The characteristic of $\lambda$-tensor identifies the type of gravitational field.  Gaussian curvature quantities enunciated in terms of curvature invariants.
\end{abstract}

% keywords can be removed
\keywords{Gravitational field \and SdS space-time \and Einstein space \and Gaussian curvature.}

\textbf{Mathematics Subject Classification 2020} 53A45 $\cdot$ 53B50 $\cdot$ 53C25 $\cdot$ 83C15

\section{Introduction}
\par Albert Einstein proposed the general theory of relativity in the year 1915. Almost one century is over after the discovery of relativity theory, but to date, it continues to yield new results in its mathematical and physical framework. On 10 April 2019, Event Horizon Telescope (EHT) released the first-ever image of a supermassive black hole in the Messier 87 galaxy. It is a milestone of astrophysics and confirms the predictions of Einstein's relativity theory. Einstein's general theory of relativity is formulated entirely in terms of differential geometry, in tensor notation, of a four-dimensional manifold combining space and time. Space-time may be described as a curved, four-dimensional mathematical structure called a semi-Riemannian manifold. In brief, space and time together comprise a curved four-dimensional non-Euclidean geometry. Einstein's theory of relativity is the geometric theory of physical space and time.
\par A Riemannian space is locally flat but globally curved. Hence Einstein used Riemannian geometry to propose general relativity, which describes how the curvature of space-time causes force and, this force recognized as gravitation. Thus the general theory of relativity is Einstein's theory of gravitation. The general relativity theory tells us that mass curves space and bends light.
\par Let $(M, g)$ be the spherically symmetric vacuum space-time. $(M, g)$ with a non-vanishing cosmological constant is called Schwarzschild-de Sitter (SdS) space-time. In other words, it is Schwarzschild space if we take cosmological constant $\Lambda=0$; it is de Sitter space interrupted by the presence of spherically symmetric mass. de Sitter space has constant negative curvature whereas Anti-de Sitter space has constant positive curvature. In cosmological studies, de Sitter space has more importance. In 1917, Einstein introduced cosmological constant. Later on, in his joint paper with de Sitter in 1932, he considered vanishing cosmological constant and told that it is the biggest blunder of his life. Nowadays positive cosmological constant is responsible for the cosmic expansion of the universe accelerated by dark energy. By recent cosmological data value of the cosmological constant is found to be $|\Lambda|<10^{-51}\,\text{m}^{-2}$.
\par The metric for SdS space-time in Schwarzschild coordinates $(t, r, \theta, \phi)$ and in relativistic units $(c=G=1)$ is given by the expression (see \cite{1})
\begin{equation}\label{e:1.1}
    g=\left(1-\frac{2m}{r}-\frac{1}{3}\Lambda r^{2}\right)dt^{2}-\left(1-\frac{2m}{r}-\frac{1}{3}\Lambda r^{2}\right)^{-1}dr^{2}-r^{2}d\Omega^{2},
\end{equation}
where $d\Omega^{2}=d\theta^{2}+\sin^{2}\theta\,d\phi^{2}$ and $m$ denotes spherically symmetric mass.

\par The SdS space-time also represents the cosmological solution, in the case $\Lambda m^{2}>1/9$, $r>0$ is time-like and $t$ is forever space-like \cite{2}. In cosmological SdS solution, the $r$-coordinate described in terms of proper time $\tau$ as follows:
\begin{equation}\label{e:1.1(a)}
    r(\tau) \propto \sinh^{2/3}\left[\left(\frac{\sqrt{3\Lambda}}{2}\right)\tau\right].
\end{equation}
The expression (\ref{e:1.1(a)}) is definitely the scale-factor in the flat $\Lambda$CDM cosmological model. Recently many geometers and cosmologists consider this scale-factor because it absolutely describes the observed rate of the cosmic expansion of our universe. Thus, the cosmological SdS solution provides an explanation of the fundamental structure of our universe. With the appearance of the cosmological SdS solution, one can interpret various events of the evolution of the universe.

\par In this paper, we follow Einstein sign convention (west coast metric sign convention, see \cite{3}) which is followed by authors like Eddington \cite{Eddington}, Schr$\ddot{\text{o}}$dinger \cite{Schrodinger}, McVittie \cite{McVittie}, Infeld and Pleba$\acute{\text{n}}$ski \cite{Infeld}, Schild \cite{Schild}, Davis \cite{Davis}, Rindler \cite{1} and many others. Recent experimental observations reveal that SdS space-time is responsible for the study of solar system effects \cite{4} such as gravitational redshift, light deflection, gravitational time delay, perihelion shift and, also to analyze the velocity of Pioneer 10 and 11 spacecraft.
\par In 2001, Wlodzimierz Borgiel \cite{5} studied the gravitational field on a pseudo-Riemannian manifold $M$ of constant curvature with a metric tensor $g$ with the aid of characteristic of $\lambda$-tensor. In 2007, the same author studied the gravitational field of the Robertson-Walker space-time \cite{6} using the characteristic of $\lambda$-tensor. In 2011, by the characteristic of $\lambda$-tensor, the same author studied the gravitational field of the Schwarzschild space-time \cite{7}. In 2013, Musavvir Ali and Zafar Ahsan \cite{8} studied the gravitational field of the Schwarzschild soliton.

\section{Invariants of the curvature tensor and nature of the gravitational field}

The Riemannian metric tensor components $g_{ij}$ determine the gravitational field. The signature of the matrix ($g_{ij}$) is $(+, -, -, -)$. The matrix ($g_{ij}$) in Schwarzschild coordinates $p\,=\,(t, r, \theta, \phi)$ is given by
\begin{equation}\label{e:1.2}
g_{ij}(p) = \left(
  \begin{array}{cccc}
    \frac{3r-6m-\Lambda r^{3}}{3r} &  &  & 0 \\
     & \frac{3r}{\Lambda r^{3}-3r+6m} &  &  \\
     &  & -r^{2} &  \\
    0 &  &  & -r^{2}\sin^{2}\theta \\
  \end{array}
\right),
\end{equation}
where $i, j=0, 1, 2, 3$. The Christoffel symbols of the second kind $\Gamma^{i}_{jk}$ determine the motion of a particle in a gravitational field and also the  intensity of gravitational field whereas the tensor $g_{ij}$ explains the role of the gravitational field potential. For real space-time the determinant of matrix $(g_{ij})$ is always negative (see \cite{9}). The determinant of the matrix $(g_{ij})$ is equal to $-r^{4}\sin^{2}\theta$. In relativistic units, gravitational field potential of SdS metric is found to be $$\Phi\approx\frac{1}{2}(g_{00}-1)=-\frac{m}{r}-\frac{1}{6}\Lambda r^{2}.$$
\par Christoffel symbols of the second kind $\Gamma^{i}_{jk}$ can be calculated using the formula
\begin{equation}\label{e:1.3}
    \Gamma^{i}_{jk} = \frac{1}{2}g^{ih}(\partial_{k}g_{jh}+\partial_{j}g_{hk}-\partial_{h}g_{kj}).
\end{equation}
There are 40 independent components. The non-vanishing components are as follows: \\
\begin{align*}
  \Gamma^{1}_{00} &= \frac{(r^{3}\Lambda-3m)(r^{3}\Lambda-3r+6m)}{9r^{3}},\,\,\,\quad\Gamma^{0}_{10}=\Gamma^{0}_{01}=\frac{r^{3}\Lambda-3m}{r(r^{3}\Lambda-3r+6m)},\\
  \Gamma^{1}_{11} &= -\frac{r^{3}\Lambda-3m}{r(r^{3}\Lambda-3r+6m)},\,\,\,\quad\Gamma^{2}_{21}=\Gamma^{2}_{12}=\frac{1}{r},\,\,\,\quad
  \Gamma^{3}_{31}=\Gamma^{3}_{13}=\frac{1}{r},\\
  \Gamma^{1}_{22} &= \frac{r^{3}\Lambda-3r+6m}{3},\,\,\,\quad\Gamma^{3}_{32}=\Gamma^{3}_{23}=\frac{\cos\theta}{\sin\theta},\\
  \Gamma^{1}_{33} &= \frac{\sin^{2}\theta (r^{3}\Lambda-3r+6m)}{3},\,\,\,\quad\Gamma^{2}_{33}=-\cos\theta\sin\theta.
\end{align*}
The curvature tensor (or Riemann tensor) can be calculated using the formula
\begin{equation}\label{e:1.4}
    R_{hijk}=\frac{1}{2}(\partial^{2}_{ij}g_{hk}+\partial^{2}_{hk}g_{ij}-\partial^{2}_{ik}g_{hj}-\partial^{2}_{hj}g_{ik})+g_{mn}(\Gamma^{m}_{ij}\Gamma^{n}_{hk} -\Gamma^{m}_{ik}\Gamma^{n}_{hj}).
\end{equation}
It has 20 independent components. The non-vanishing components of the curvature tensor $R_{hijk}$ are
\begin{align*}
  R_{1010}(p) &= \frac{r^{3}\Lambda+6m}{3r^{3}},\,\,\,\quad R_{2020}(p)=-\frac{(r^{3}\Lambda-3m)(r^{3}\Lambda-3r+6m)}{9r^2},\\
  R_{2121}(p) &= \frac{r^{3}\Lambda-3m}{r^{3}\Lambda-3r+6m},\,\,\,\quad R_{3030}(p)=-\frac{\sin^{2}\theta (r^{3}\Lambda-3m)(r^{3}\Lambda-3r+6m)}{9r^{2}},\\
  R_{3131}(p) &= \frac{\sin^{2}\theta (r^{3}\Lambda-3m)}{r^{3}\Lambda-3r+6m},\,\,\,\quad R_{3232}(p)=-\frac{r\sin^{2}\theta (r^{3}\Lambda+6m)}{3}.
\end{align*}
The Ricci tensor can be expressed as
\begin{equation*}
    R_{ij}=R^{l}_{ijl},
\end{equation*}
and we found that $R_{ij}=\Lambda g_{ij}$.\\
The scalar curvature of the space is an invariant and it can be calculated using the expression
\begin{equation*}
    R=g^{ij}R_{ij}.
\end{equation*}
For our space-time, it is found to be $4\Lambda$. Hence we conclude that scalar curvature is proportional to cosmological constant. So that SdS space-time is an Einstein space of constant scalar curvature. \\
Kretschmann scalar is given by
\begin{equation*}
    R^{hijk}R_{hijk}=\frac{8(r^{6}\Lambda^{2}+18m^{2})}{3r^{6}}.
\end{equation*}
Einstein tensor $G_{ij}$ can be expressed as
\begin{equation*}
    G_{ij}=R_{ij}-\frac{1}{2}g_{ij}R.
\end{equation*}
It is found that $G_{ij}=-\Lambda g_{ij}$. Hence Einstein's Field Equation (EFE) is given by
\begin{equation*}
    G_{ij}+\Lambda g_{ij}=0.
\end{equation*}
Therefore EFE can't reduce to $G_{ij}=0$ even in the absence of sources.
\par Next, we calculate the non-vanishing components of Weyl conformal curvature tensor using the expression
\begin{equation*}
    C_{hijk}=R_{hijk}+g_{j[h}R_{i]k}-g_{k[h}R_{i]j}-\frac{1}{3}g_{j[h}g_{i]k}R,
\end{equation*}
where the square brackets indicate the antisymmetric property over the indices contained in them:
\begin{equation*}
    P_{[ij]}=\frac{1}{2}(P_{ij}-P_{ji}).
\end{equation*}
The Weyl conformal curvature tensor obtained from curvature tensor, Ricci tensor and scalar curvature. It satisfies the symmetry properties as curvature tensor but it is traceless. When contracted on the pair of indices $hk$ or $ij$, the value of the tensor becomes zero. It has 10 independent components. Among them, the non-zero components of the Weyl tensor are as follows:
\begin{align*}
  C_{1010}(p) &= \frac{2m}{r^{3}},\,\,\,\,\quad C_{2020}(p)=\frac{m(r^{3}\Lambda-3r+6m)}{3r^{2}},\\
  C_{3030}(p) &= \frac{m\sin^{2}\theta(r^{3}\Lambda-3r+6m)}{3r^{2}},\,\,\,\quad C_{1212}(p)=-\frac{3m}{r^{3}\Lambda-3r+6m},\\
  C_{3131}(p) &= -\frac{3m\sin^{2}\theta}{r^{3}\Lambda-3r+6m},\,\,\,\quad C_{2323}(p)=-2mr\sin^{2}\theta.
\end{align*}
Since some components of Weyl conformal curvature tensor at an arbitrary point of SdS space-time are non-zero, SdS space-time is not conformally flat. However both de Sitter space and Anti-de Sitter space are conformally flat.
\par To discuss the Riemann tensor and the bivector-tensors, it is convenient to introduce six-dimensional formalism in the pseudo-Euclidean space (see \cite{10}). The rule for switching into the six-dimensional formalism is the following:
\begin{eqnarray*}
  hi &:& 23\,\,\,31\,\,\,12\,\,\,10\,\,\,20\,\,\,30 \\
  I &:& \,\,\,1\,\,\,\,\,\,2\,\,\,\,\,\,3\,\,\,\,\,\,4\,\,\,\,\,\,5\,\,\,\,\,\,6
\end{eqnarray*}
\par According to the above scheme, a bivector-tensor corresponds to the symmetric tensor in the six-dimensional pseudo-Euclidean space. The components of the curvature tensor $R_{hijk}$ change over to the components of the symmetric tensor $R_{IJ}\,(I,\,J=1,\,2,\,3,\,4,\,5,\,6)$ in the six-dimensional pseudo-Euclidean space provided each of the suffix pairs $hi, jk$ are re-identified according to the scheme of six-dimensional formalism. The bivector-tensor is defined as
\begin{equation*}
    g_{IJ}=g_{hijk}=g_{hj}g_{ik}-g_{hk}g_{ij},
\end{equation*}
where $g_{ij}$ are the components of the metric tensor at an arbitrary point of SdS space-time. The tensor $g_{IJ}$ is symmetric and non-singular. It is of signature $(+,\,+,\,+,\,-,\,-,\,-)$. The suffix pairs $hi, jk$ are skew-symmetric in the definition of bivector-tensor.\\
The non-vanishing components of tensor $g_{IJ}$ are
\begin{align*}
  g_{11}(p) &= r^{4}\sin^{2}\theta,\,\,\,\quad g_{22}(p)=\frac{3r^{3}\sin^{2}\theta}{3r-r^{3}\Lambda-6m},\,\,\,\quad g_{33}(p)=\frac{3r^{3}}{3r-r^{3}\Lambda-6m} \\
  g_{44}(p) &= -1,\,\,\,\quad g_{55}(p)=-\frac{r(3r-r^3\Lambda-6m)}{3},\,\,\,\quad g_{66}(p)=-\frac{r\sin^{2}\theta(3r-r^3\Lambda-6m)}{3}.
\end{align*}
The non-vanishing components of the symmetric tensor $R_{IJ}$ are
\begin{align*}
  R_{11}(p) &= -\frac{r\sin^{2}\theta(r^3\Lambda+6m)}{3},\,\,\,\quad R_{22}(p)=\frac{\sin^{2}\theta(r^3\Lambda-3m)}{r^3\Lambda-3r+6m},\\
  R_{33}(p) &= \frac{r^3\Lambda-3m}{r^3\Lambda-3r+6m},\,\,\,\quad R_{44}(p)=\frac{r^3\Lambda+6m}{3r^{3}},\\
  R_{55}(p) &= -\frac{(r^3\Lambda-3m)(r^3\Lambda-3r+6m)}{9r^{2}},\,\,\,\quad R_{66}(p)=-\frac{\sin^{2}\theta(r^3\Lambda-3m)(r^3\Lambda-3r+6m)}{9r^{2}}.
\end{align*}
$R_{IJ}-\lambda g_{IJ}$ is known as $\lambda$-tensor. The roots of the equation $|R_{IJ}(p)-\lambda g_{IJ}(p)|=0$ are called curvature invariants. For our SdS space-time, invariants of the curvature tensor (see \cite{9}) are as follows:
\begin{align}
\label{e:1.5}  \lambda_{1}(r)  &= \lambda_{4}(r)=-\frac{r^3\Lambda+6m}{3r^{3}}, \\
\label{e:1.6}  \lambda_{2}(r) &= \lambda_{3}(r)=\lambda_{5}(r)=\lambda_{6}(r)=-\frac{r^3\Lambda-3m}{3r^{3}}.
\end{align}
For any of the above curvature invariants, determinant of the $\lambda$-matrix $R_{IJ}(p)-\lambda g_{IJ}(p)$ is zero. The canonical form of the curvature tensor is given by
\begin{center}
$R_{I'J'}=\left(
  \begin{array}{cccccc}
    -\frac{r^3\Lambda+6m}{3r^{3}} &  &  &  &  & 0 \\
     & -\frac{r^3\Lambda-3m}{3r^3} &  &  &  &  \\
     &  & -\frac{r^3\Lambda-3m}{3r^3} &  &  &  \\
     &  &  & \frac{r^3\Lambda+6m}{3r^{3}} &  &  \\
     &  &  &  & \frac{r^3\Lambda-3m}{3r^3} &  \\
    0 &  &  &  &  & \frac{r^3\Lambda-3m}{3r^3} \\
  \end{array}
\right).$
\end{center}
Also we have
\begin{center}
$g_{I'J'}=\left(
  \begin{array}{cccccc}
    1 &  &  &  &  & 0 \\
     & 1 &  &  &  &  \\
     &  & 1 &  &  &  \\
     &  &  & -1 &  &  \\
     &  &  &  & -1 &  \\
    0 &  &  &  &  & -1 \\
  \end{array}
\right).$
\end{center}
\par Hence we conclude from the characteristic of $\lambda$-tensor $R_{IJ}-\lambda g_{IJ}$ that the gravitational field of the SdS non-singular space-time is the type $G_{1}[(1111)(11)]$ in the Segre characteristic (Ref. \cite{11}).
\par According to the algebraic structure of Riemann tensor, the presented SdS space-time is of Type I in the Petrov's classification of spaces defining gravitational fields (Ref. \cite{12}). Also, we noted that SdS space-time is not flat even in the absence of spherically symmetric mass due to consideration of the non-zero cosmological constant in EFE.
\par Now suppose $\theta=0$ or $\theta=\pi$. Since $\theta$ is constant we have $d\theta=0$. Then the SdS metric modified to the following form
\begin{equation}\label{e:1.7}
    g'=\left(1-\frac{2m}{r}-\frac{1}{3}\Lambda r^{2}\right)dt^{2}-\left(1-\frac{2m}{r}-\frac{1}{3}\Lambda r^{2}\right)^{-1}dr^{2}.
\end{equation}
The matrix ($g'_{ij}$) in coordinates $p'\,=\,(t, r)$ is given by
\begin{equation}\label{e:1.8}
   g'_{ij}(p')=\left(
  \begin{array}{cc}
    -\frac{r^3\Lambda-3r+6m}{3r} & 0 \\
    0 & \frac{3r}{r^3\Lambda-3r+6m} \\
  \end{array}
\right),
\end{equation}
where $i, j=0, 1$. The determinant of the matrix ($g'_{ij}$) is $-1$. The non-zero components of the Christoffel symbols of second kind and the curvature tensor for the metric (\ref{e:1.7}) are as follows:
\begin{align*}
  \Gamma'^{1}_{00} &= \frac{(r^3\Lambda-3m)(r^3\Lambda-3r+6m)}{9r^{3}},\quad\Gamma'^{0}_{01}=\Gamma'^{0}_{10}=\frac{r^3\Lambda-3m}{r(r^3\Lambda-3r+6m)}, \\
  \Gamma'^{1}_{11} &= -\frac{r^3\Lambda-3m}{r(r^3\Lambda-3r+6m)},\,\,\quad\quad R'_{1010}(p')=\frac{r^3\Lambda+6m}{3r^{3}}.
\end{align*}
A two-dimensional surface is degenerated by the hypersurface $H'_{0}$ or $H'_{\pi}$. Every point of two-dimensional surface is isotropic. The Gaussian curvature of the hypersurface $H'_{0}$ or $H'_{\pi}$ can be expressed as
\begin{equation}\label{e:1.9}
    K'(p')=\frac{R'_{1010}(p')}{\left|
  \begin{array}{cc}
    -\frac{r^3\Lambda-3r+6m}{3r} & 0 \\
    0 & \frac{3r}{r^3\Lambda-3r+6m} \\
  \end{array}
\right|}=-\frac{r^3\Lambda+6m}{3r^{3}}.
\end{equation}
Now we consider the case in which $\theta\in(0, \pi)$ and $\phi=0$. Since $\phi$ is constant we have $d\phi=0$. Then the SdS metric modified to the following form
\begin{equation}\label{e:1.10}
    g''=\left(1-\frac{2m}{r}-\frac{1}{3}\Lambda r^{2}\right)dt^{2}-\left(1-\frac{2m}{r}-\frac{1}{3}\Lambda r^{2}\right)^{-1}dr^{2}-r^{2}d\theta^{2}.
\end{equation}
The matrix ($g''_{ij}$) in coordinates $p''\,=\,(t, r, \theta)$ is given by
\begin{equation}\label{e:1.11}
    g''_{ij}(p'')=\left(
                    \begin{array}{ccc}
                     -\frac{r^3\Lambda-3r+6m}{3r}  &  & 0 \\
                       & \frac{3r}{r^3\Lambda-3r+6m} &  \\
                      0 &  & -r^{2} \\
                    \end{array}
                  \right),
\end{equation}
where $i, j=0, 1, 2$. The determinant of the matrix ($g''_{ij}$) is $r^{2}$. The non-zero components of the Christoffel symbols of second kind for the metric (\ref{e:1.10}) are found to be
\begin{align*}
  \Gamma''^{1}_{00} &= \frac{(r^3\Lambda-3m)(r^3\Lambda-3r+6m)}{9r^{3}},\,\,\,\quad\Gamma''^{0}_{10}=\Gamma''^{0}_{01}=\frac{r^3\Lambda-3m}{r(r^3\Lambda-3r+6m)}, \\
  \Gamma''^{1}_{11} &= -\frac{r^3\Lambda-3m}{r(r^3\Lambda-3r+6m)},\,\,\,\quad\Gamma''^{2}_{12}=\Gamma''^{2}_{21}=\frac{1}{r},\,\,\,\quad\Gamma''^{1}_{22}=\frac{r^3\Lambda-3r+6m}{3}.
\end{align*}
The non-vanishing components of the curvature tensor for the metric (\ref{e:1.10}) are obtained as follows:
\begin{align*}
  R''_{1010}(p'') &= \frac{r^3\Lambda+6m}{3r^{3}},\,\,\,\quad R''_{2020}(p'')=-\frac{(r^3\Lambda-3m)(r^3\Lambda-3r+6m)}{9r^{2}}, \\
  R''_{2121}(p'') &= \frac{r^3\Lambda-3m}{r^3\Lambda-3r+6m}.
\end{align*}
So in three-dimensional space, the Riemann tensor has three non-vanishing independent components. Hence three quantities of Gaussian curvature determine the curvature of three-dimensional space at each point $p''$ and are expressed as
\begin{align}
\label{e:1.12}  K''_{\theta}(p'') &= \frac{R''_{1010}(p'')}{\left|
                                                 \begin{array}{cc}
                                                  -\frac{r^3\Lambda-3r+6m}{3r}  & 0 \\
                                                   0 & \frac{3r}{r^3\Lambda-3r+6m} \\
                                                 \end{array}
                                               \right|}=-\frac{r^3\Lambda+6m}{3r^{3}}, \\
\label{e:1.13}  K''_{r}(p'') &= \frac{R''_{2020}(p'')}{\left|
                                           \begin{array}{cc}
                                             -\frac{r^3\Lambda-3r+6m}{3r} & 0 \\
                                             0 & -r^{2} \\
                                           \end{array}
                                         \right|}=-\frac{r^3\Lambda-3m}{3r^{3}}, \\
\label{e:1.14}  K''_{t}(p'') &= \frac{R''_{2121}(p'')}{\left|
                                            \begin{array}{cc}
                                             \frac{3r}{r^3\Lambda-3r+6m}  & 0 \\
                                              0 & -r^{2} \\
                                            \end{array}
                                          \right|}=-\frac{r^3\Lambda-3m}{3r^{3}}.
\end{align}
Now we point out that the quantities (\ref{e:1.9}) and (\ref{e:1.12}) are equal to the curvature invariants (\ref{e:1.5}) and the quantities (\ref{e:1.13}) and (\ref{e:1.14}) are equal to the curvature invariants (\ref{e:1.6}).
\par Finally, we identified that six quantities of Gaussian curvature $K_{t, r}(p), K_{t, \theta}(p), K_{t, \phi}(p)$, $K_{r, \theta}(p),  K_{r, \phi}(p), K_{\theta, \phi}(p)$ determine the curvature of SdS space-time $(M, g)$ at each point $p\in M$. Six Gaussian curvature quantities are listed as follows:
\begin{equation}\label{e:1.15}
\left.\begin{aligned}
K_{t, r}(p) &= K'(p')=K''_{\theta}(p'')=\lambda_{1}(r), \\
K_{t, \theta}(p) &= K''_{r}(p'')=\lambda_{2}(r), \\
K_{t, \phi}(p) &= K''_{t}(p'')=\lambda_{3}(r), \\
K_{r, \theta}(p) &= \lambda_{5}(r), \\
K_{r, \phi}(p) &= \lambda_{6}(r), \\
K_{\theta, \phi}(p) &= \lambda_{4}(r).
\end{aligned}\right\rbrace
\end{equation}
\par
\begin{figure}
    \centering
	\includegraphics[width=4.5 in]{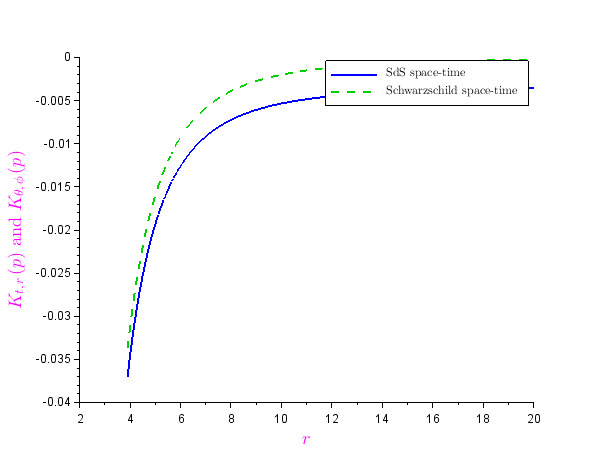}
	\caption{Effect of cosmological constant on Gaussian curvature quantities $K_{t, r}(p)$ and $K_{\theta, \phi}(p)$.}
	\label{f:1.1}
\end{figure}

\begin{figure}
    \centering
    \includegraphics[width=4.5 in]{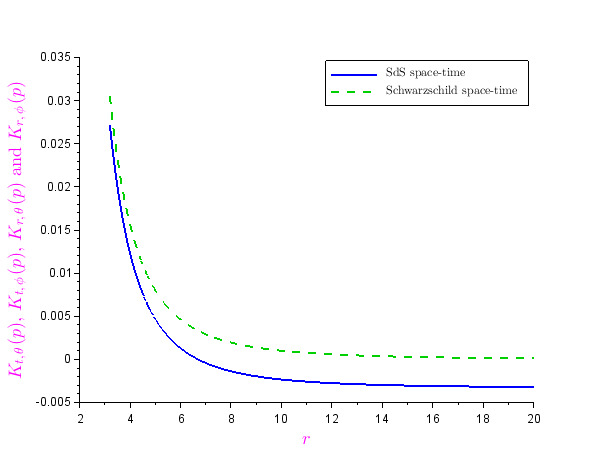}
	\caption{Effect of cosmological constant on Gaussian curvature quantities $K_{t, \theta}(p), K_{t, \phi}(p), K_{r, \theta}(p)$ and $K_{r, \phi}(p)$.}
	\label{f:1.2}
\end{figure}

 We have studied the effect of non-zero cosmological constant on Gaussian curvature quantities through the Figs.\,\ref{f:1.1} and \ref{f:1.2}. $r$ is plotted versus $K_{t, r}(p)$ and $K_{\theta, \phi}(p)$ in Fig.\,\ref{f:1.1} and it is plotted versus $K_{t, \theta}(p), K_{t, \phi}(p), K_{r, \theta}(p)$ and $K_{r, \phi}(p)$ in Fig.\,\ref{f:1.2}. In both plots we consider $m=1$ and $\Lambda=0.01$.

\par From Fig.\,\ref{f:1.1} it is observed that at every point the Gaussian curvature quantities $K_{t, r}(p)$ and $K_{\theta, \phi}(p)$ of SdS space-time are less than that of Schwarzschild space-time and similarly from Fig.\,\ref{f:1.2} we found that at every point the Gaussian curvature quantities $K_{t, \theta}(p), K_{t, \phi}(p), K_{r, \theta}(p)$ and $K_{r, \phi}(p)$ of SdS space-time are also less than that of Schwarzschild space-time. Further, from both the plots, it should be noted that as $r\rightarrow\infty$, the curvature of Schwarzschild space-time tends to zero and, hence it becomes flat, whereas curvature of SdS space-time never tends to zero. Therefore, SdS space-time never becomes flat at a very distant point.

\par For large $r$ and $\Lambda>0$, SdS space-time approximates de Sitter space. In this case, all curvature invariants are negative. All six quantities of Gaussian curvature are negative at all points of de Sitter space. In the gravitational field of de Sitter space, the sum of the angles of a small triangle in a plane which passes through the centre is less than 180 degrees. Hence every point of de Sitter space is a hyperbolic point. For large $r$ and $\Lambda<0$, SdS space-time approximates Anti-de Sitter space. In the gravitational field of Anti-de Sitter space, the sum of the angles of a small triangle in a plane which passes through the centre is greater than 180 degrees. Hence every point of Anti-de Sitter space is an elliptic point. For $\Lambda=0$, SdS space-time reduces to Schwarzschild space-time. Further, we pointed out that for authors who consider the east coast metric sign convention (see \cite{3}), that is $(-, +, +, +)$, every point of de Sitter space is an elliptic point and, that of Anti-de Sitter space is a hyperbolic point.

\par In the SdS metric (\ref{e:1.1}) with $m=0$ and $\Lambda>0$, The quantity $g_{00}$ vanishes and $g_{11}$ goes to infinity at $r=\sqrt{\frac{3}{\Lambda}}$. Thus we may conclude that the SdS space-time metric has a singularity. The centre of the field $(r=0)$ is a location of the genuine curvature singularity of the metric. The Gaussian curvature of the entire space-time described by the SdS metric becomes infinite as $r\rightarrow0$. The static radius of the SdS black hole is the hypersurface where cosmic repulsion balances its gravitational attraction. Similar to the Schwarzschild metric inside its horizon, the SdS metric inside the black hole horizon also becomes non-static.

\par
\begin{figure}
    \centering
    \includegraphics[width=4.5 in]{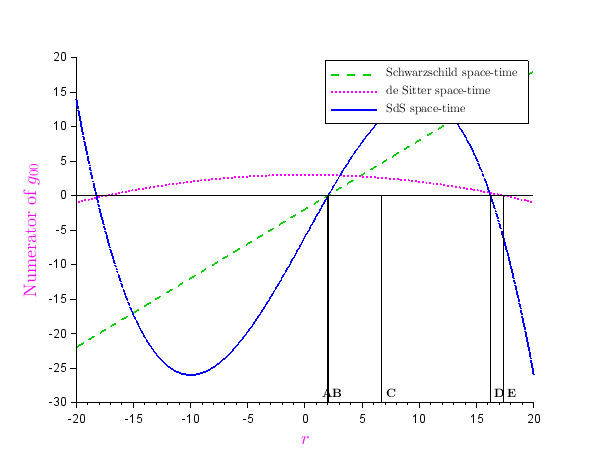}
    \caption{Coordinate singularities.}
    \label{f:1.3}
\end{figure}

 Coordinate singularities of Schwarzschild space-time, de Sitter space-time, and SdS space-time are illustrated graphically in Fig.\,\ref{f:1.3} wherein $r$ is plotted versus numerator of $g_{00}$ for $m=1$ and $\Lambda=0.01$.

\par
\begin{figure}
    \centering
    \includegraphics[width=4.5 in]{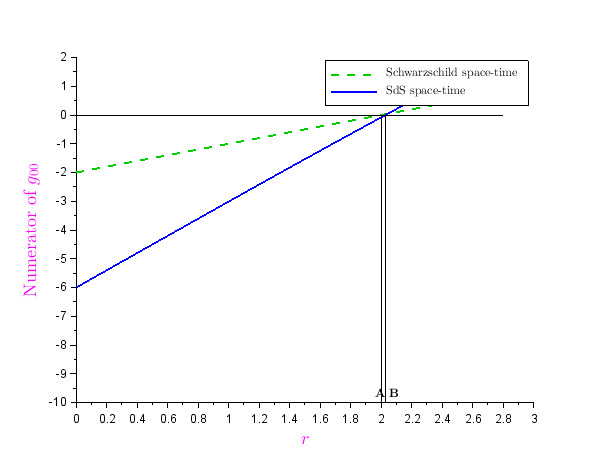}
    \caption{Schwarzschild horizon and SdS black hole horizon.}
    \label{f:1.4}
\end{figure}

 For SdS space-time the black hole horizon, static radius, and cosmological horizon are located at \textbf{B}, \textbf{C}, and \textbf{D}, respectively. Schwarzschild horizon and de Sitter horizon are located at \textbf{A} and \textbf{E}, respectively, as shown in Fig.\,\ref{f:1.3}. From Fig.\,\ref{f:1.3} it is clear that Schwarzschild horizon and SdS black hole horizon are very close to each other, which is also interpreted in Fig.\,\ref{f:1.4}.

\par
\begin{figure}
  \centering
  \includegraphics[width=4.5 in]{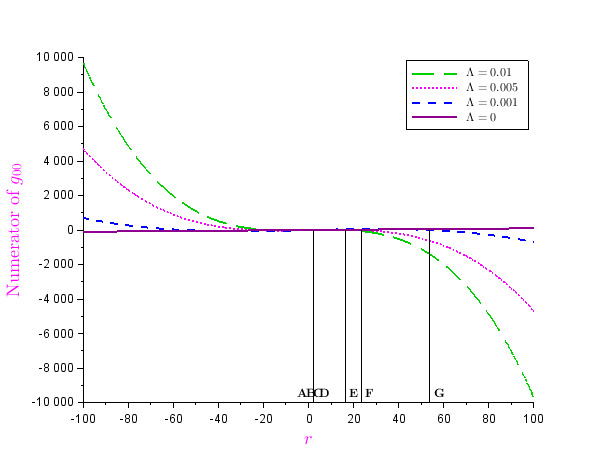}
  \caption{Variation of graph of SdS space-time with cosmological constant.}
  \label{f:1.5}
\end{figure}

 Next, we have plotted the graph of SdS space-time for different values of the cosmological constant $\Lambda$. $r$ is plotted versus numerator of $g_{00}$ in Fig.\,\ref{f:1.5}. Here we consider $\Lambda=0.01, \Lambda=0.005, \Lambda=0.001$ and pure Schwarzschild case with $\Lambda=0$. From Fig.\,\ref{f:1.5}, we found that as the value of $\Lambda$ decreases and tends to zero, SdS space-time approaches Schwarzschild space-time. With decreasing value of $\Lambda$, the black hole and cosmological horizons are going away from each other, that is, a separation between the black hole and cosmological horizons increases as $\Lambda$ decreases.
\par
\begin{figure}
    \centering
    \includegraphics[width=4.5 in]{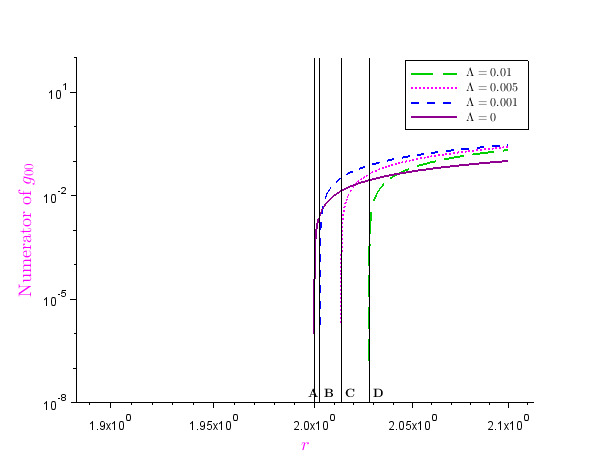}
    \caption{Variation of position of SdS black hole horizon with cosmological constant.}
    \label{f:1.6}
\end{figure}

 For SdS space-time with $\Lambda=0.01$, the black hole and cosmological horizons are located at \textbf{D} and \textbf{E}, respectively; with $\Lambda=0.005$, they are located at \textbf{C} and \textbf{F}, respectively; with $\Lambda=0.001$, they are located at \textbf{B} and \textbf{G}, respectively. For pure Schwarzschild case with $\Lambda=0$, the event horizon is located at \textbf{A} as shown in Fig.\,\ref{f:1.5}. In Fig.\,\ref{f:1.5}, SdS black hole horizons for different values of $\Lambda$ and Schwarzschild horizon are very close to each other. It is also shown in Fig.\,\ref{f:1.6}.
\par From Fig.\,\ref{f:1.6}, we noticed that as $\Lambda$ tends to zero, the SdS black hole horizon approaches the Schwarzschild event horizon.

\newpage

\section{Conclusions and perspectives}

\par We have investigated the gravitational field of the SdS Space-time. We found that in four orientations, we have the Gaussian curvature equal to $-\frac{r^3\Lambda-3m}{3r^{3}}$ and in the remaining two orientations we have the Gaussian curvature equal to $-\frac{r^3\Lambda+6m}{3r^{3}}$. Hence we conclude that every point of SdS space-time is not isotropic. But in the absence of spherically symmetric mass, that is, in the case of de Sitter space or Anti-de Sitter space we have the Gaussian curvature equal to $-\frac{\Lambda}{3}$ in all orientations. So every point of de Sitter space and that of Anti-de Sitter space is isotropic. Therefore we have established that both de Sitter space and Anti-de Sitter space are completely isotropic, that is,
\begin{align*}
R_{hijk} &= K(g_{hj}g_{ik}-g_{hk}g_{ij}), \\
R'_{hijk} &= K(g'_{hj}g'_{ik}-g'_{hk}g'_{ij}), \\
R''_{hijk} &= K(g''_{hj}g''_{ik}-g''_{hk}g''_{ij}),
\end{align*}
where $K=-\frac{\Lambda}{3}$ is the Gaussian curvature.
\par We have obtained canonical form of the tensor $R_{IJ}$ in the pseudo-Euclidean space by six-dimensional formalism with respect to orthonormal basis $e^{1}, e^{2}, e^{3}, e^{4}, e^{5}, e^{6}$. The components of the canonical form of the tensor $R_{IJ}$ describe the curvature of the space-time; in turn, it leads to the study of nature of the gravitational field. The curvature invariants in orthonormal directions $e^{i}\,(i=1, \ldots, 6)$ are equal to one another for de Sitter space, Anti-de Sitter space. So we may conclude that the gravitational field of de Sitter space and Anti-de Sitter space is constant.
\par We need to analyze the study of the gravitational field of rotating and electrically charged SdS black holes. It may be a possible extension of this work, and we have planned to study it in the future. So we leave the analysis of axially symmetric vacuum space-time with non-vanishing cosmological constant as a perspective.
\par In order to study our universe, our space-time has to be considered with a dynamic matter and radiation description. So, we have planned to study cosmological SdS solution and, its geometric and physical properties in our universe, in the future work. Further, the study of the thermodynamic properties of black holes such as temperature and entropy, with multiple horizons, is in progress.

\section*{Acknowledgments}

We wish to thank Karnataka Science and Technology Promotion Society (KSTePS), Department of Science and Technology (DST), Govt. of Karnataka
(Grant No. OTH-04:\,2018-19) for the support of this work. We are grateful to the DST, Govt. of India, for providing financial assistance (Grant No. SR/FST/MS-I/2018/23(C)) to the Department of Mathematics, Kuvempu University, Karnataka, India under FIST Program.

\bibliographystyle{unsrt}

\end{document}